\documentclass[aps,prapplied,reprint,groupedaddress,superscriptaddress,longbibliography,float]{revtex4-2}
\usepackage{endnotes}
\usepackage{amsfonts}
\usepackage{amsmath}
\usepackage{bbm}
\usepackage{mathrsfs}
\usepackage{graphics,graphicx,epsfig,bm,amsmath,amsthm,amssymb}
\usepackage{bm}
\usepackage{bbm}
\usepackage{longtable}
\usepackage{multirow}
\usepackage{array}
\usepackage{color}
\usepackage[usenames,dvipsnames]{xcolor}
\usepackage[super]{nth}
\usepackage{lipsum}
\usepackage{mathrsfs}
\usepackage[english]{babel}
  
\usepackage{float}
\usepackage[a4paper,colorlinks=true,
linkcolor=blue,citecolor=blue,
pdfauthor={ },
pdftitle={ },
pdfsubject={ },
pdfkeywords={ }]{hyperref}

\begin{document}
\title{Observation of uniaxial strain tuned spin cycloid in a freestanding BiFeO$_3$ film}

\author{Zhe Ding}
\email{These authors contributed equally to this work.}
\affiliation{CAS Key Laboratory of Microscale Magnetic Resonance and School of Physical Sciences, University of Science and Technology of China, Hefei 230026, China}
\affiliation{CAS Center for Excellence in Quantum Information and Quantum Physics, University of Science and Technology of China, Hefei 230026, China}
\affiliation{Hefei National Laboratory, University of Science and Technology of China, Hefei 230088, China}

\author{Yumeng Sun}
\email{These authors contributed equally to this work.}
\affiliation{CAS Key Laboratory of Microscale Magnetic Resonance and School of Physical Sciences, University of Science and Technology of China, Hefei 230026, China}
\affiliation{CAS Center for Excellence in Quantum Information and Quantum Physics, University of Science and Technology of China, Hefei 230026, China}
\affiliation{Hefei National Laboratory, University of Science and Technology of China, Hefei 230088, China}

\author{Ningchong Zheng}
\email{These authors contributed equally to this work.}
\affiliation{National Laboratory of Solid State Microstructures, Nanjing University, Nanjing 210093, China}
\affiliation{Jiangsu Key Laboratory of Artificial Functional Materials, Nanjing University, Nanjing 210093, China}
\affiliation{College of Engineering and Applied Science, Nanjing University, Nanjing 210093, China}
\affiliation{Collaborative Innovation Center of Advanced Microstructures, Nanjing University, Nanjing 210093, China}

\author{Xingyue Ma}
\email{These authors contributed equally to this work.}
\affiliation{National Laboratory of Solid State Microstructures, Nanjing University, Nanjing 210093, China}
\affiliation{Jiangsu Key Laboratory of Artificial Functional Materials, Nanjing University, Nanjing 210093, China}
\affiliation{College of Engineering and Applied Science, Nanjing University, Nanjing 210093, China}
\affiliation{Collaborative Innovation Center of Advanced Microstructures, Nanjing University, Nanjing 210093, China}

\author{Mengqi Wang}
\affiliation{CAS Key Laboratory of Microscale Magnetic Resonance and School of Physical Sciences, University of Science and Technology of China, Hefei 230026, China}
\affiliation{CAS Center for Excellence in Quantum Information and Quantum Physics, University of Science and Technology of China, Hefei 230026, China}
\affiliation{Hefei National Laboratory, University of Science and Technology of China, Hefei 230088, China}

\author{Yipeng Zang}
\affiliation{National Laboratory of Solid State Microstructures, Nanjing University, Nanjing 210093, China}
\affiliation{Jiangsu Key Laboratory of Artificial Functional Materials, Nanjing University, Nanjing 210093, China}
\affiliation{College of Engineering and Applied Science, Nanjing University, Nanjing 210093, China}
\affiliation{Collaborative Innovation Center of Advanced Microstructures, Nanjing University, Nanjing 210093, China}

\author{Pei Yu}
\affiliation{CAS Key Laboratory of Microscale Magnetic Resonance and School of Physical Sciences, University of Science and Technology of China, Hefei 230026, China}
\affiliation{CAS Center for Excellence in Quantum Information and Quantum Physics, University of Science and Technology of China, Hefei 230026, China}
\affiliation{Hefei National Laboratory, University of Science and Technology of China, Hefei 230088, China}


\author{Pengfei Wang}
\affiliation{CAS Key Laboratory of Microscale Magnetic Resonance and School of Physical Sciences, University of Science and Technology of China, Hefei 230026, China}
\affiliation{CAS Center for Excellence in Quantum Information and Quantum Physics, University of Science and Technology of China, Hefei 230026, China}
\affiliation{Hefei National Laboratory, University of Science and Technology of China, Hefei 230088, China}

\author{Ya Wang}
\affiliation{CAS Key Laboratory of Microscale Magnetic Resonance and School of Physical Sciences, University of Science and Technology of China, Hefei 230026, China}
\affiliation{CAS Center for Excellence in Quantum Information and Quantum Physics, University of Science and Technology of China, Hefei 230026, China}
\affiliation{Hefei National Laboratory, University of Science and Technology of China, Hefei 230088, China}

\author{Yurong Yang}
\email{yangyr@nju.edu.cn}
\affiliation{National Laboratory of Solid State Microstructures, Nanjing University, Nanjing 210093, China}
\affiliation{Jiangsu Key Laboratory of Artificial Functional Materials, Nanjing University, Nanjing 210093, China}
\affiliation{College of Engineering and Applied Science, Nanjing University, Nanjing 210093, China}
\affiliation{Collaborative Innovation Center of Advanced Microstructures, Nanjing University, Nanjing 210093, China}

\author{Yuefeng Nie}
\email{ynie@nju.edu.cn}
\affiliation{National Laboratory of Solid State Microstructures, Nanjing University, Nanjing 210093, China}
\affiliation{Jiangsu Key Laboratory of Artificial Functional Materials, Nanjing University, Nanjing 210093, China}
\affiliation{College of Engineering and Applied Science, Nanjing University, Nanjing 210093, China}
\affiliation{Collaborative Innovation Center of Advanced Microstructures, Nanjing University, Nanjing 210093, China}

\author{Fazhan Shi}
\affiliation{CAS Key Laboratory of Microscale Magnetic Resonance and School of Physical Sciences, University of Science and Technology of China, Hefei 230026, China}
\affiliation{CAS Center for Excellence in Quantum Information and Quantum Physics, University of Science and Technology of China, Hefei 230026, China}
\affiliation{Hefei National Laboratory, University of Science and Technology of China, Hefei 230088, China}
\affiliation{School of Biomedical Engineering and Suzhou Institute for Advanced Research, University of Science and Technology of China, Suzhou 215123, China}

\author{Jiangfeng Du}
\email{djf@ustc.edu.cn}
\affiliation{CAS Key Laboratory of Microscale Magnetic Resonance and School of Physical Sciences, University of Science and Technology of China, Hefei 230026, China}
\affiliation{CAS Center for Excellence in Quantum Information and Quantum Physics, University of Science and Technology of China, Hefei 230026, China}
\affiliation{Hefei National Laboratory, University of Science and Technology of China, Hefei 230088, China}

\begin{abstract}
Non-collinear spin order that breaks space inversion symmetry and allows efficient electric-field control of magnetism makes BiFeO$_3$ a promising candidate for applications in low-power spintronic devices\cite{kimuraMagneticControlFerroelectric2003,lottermoserMagneticPhaseControl2004,cheongMultiferroicsMagneticTwist2007a,heronDeterministicSwitchingFerromagnetism2014a}. Epitaxial strain effects have been intensively studied and exhibit significant modulation of the magnetic order in BiFeO$_3$\cite{sandoCraftingMagnonicSpintronic2013, haykalAntiferromagneticTexturesBiFeO2020}, but tuning its spin structure with continuously varied uniaxial strain is still lacking up to date. Here, we apply \emph{in situ} uniaxial strain to a freestanding BiFeO$_3$ film and use scanning NV microscope to image the nanoscale magnetic order in real-space. The strain is continuously increased from 0\% to 1.5\% and four images under different strains are acquired during this period. The images show that the spin cycloid tilts by $\sim 12.6^\circ$ when strain approaches 1.5\%. A first principle calculation has been processed to show that the tilting is energetically favorable under such strain. Our \emph{in situ} strain applying method in combination with scanning NV microscope real-space imaging ability paves a new way in studying the coupling between magnetic order and strain in BiFeO$_3$ films. 
\end{abstract}

\maketitle

Antiferromagnetic material is robust against external magnetic field disturb, has super-fast spin dynamics and possesses large magneto-transport effects. Due to the merits above, antiferromagnetic materials have important application in spintronics and other magnetism-based techniques \cite{baltzAntiferromagneticSpintronics2018}.  Although it is a promising material, because of its anti-parallel spin configuration which leads to zero stray-field, antiferromagnetic material cannot be well studied by normal near-field imaging techniques   \cite{grossRealspaceImagingNoncollinear2017}. Non-collinear antiferromagnetic perovskite compound bismuth ferrite (BiFeO$_3$, BFO) is the only magnetoelectric multiferroic material under room temperature. Since the non-collinear spin cycloid breaks spatial inversion symmetry, it can be controlled by external electric field and thus costs much less energy comparing to ordinary ferromagnetic devices \cite{grossRealspaceImagingNoncollinear2017}. BFO owns spin cycloid because of the Dzyaloshinskii–Moriya interaction (DMI), such cycloid induces an effective magnetization which is too weak to detect with normal methods such as MFM\cite{hartmannMagneticForceMicroscopy1999} and PEEM\cite{locatelliRecentAdvancesChemical2008}. At the same time, since BFO has $\sim 2.7$eV bandgap\cite{sandoLargeElastoopticEffect2016}, sp-STM also lacks the ability to perform imaging. Scanning NV microscopy (SNVM) is an emergent real-space scanning method with nanoscale spatial resolution and $\mu \rm{T} /\sqrt{\rm{Hz}}$ magnetic sensitivity\cite{taylorHighsensitivityDiamondMagnetometer2008, maletinskyRobustScanningDiamond2012}. People have utilized SNVM to study the magnetic structure of BFO epitaxial films at nanoscale \cite{ grossRealspaceImagingNoncollinear2017, haykalAntiferromagneticTexturesBiFeO2020, chauleauElectricAntiferromagneticChiral2020}. 

\begin{figure*}
\centering
\includegraphics[width=0.8\linewidth]{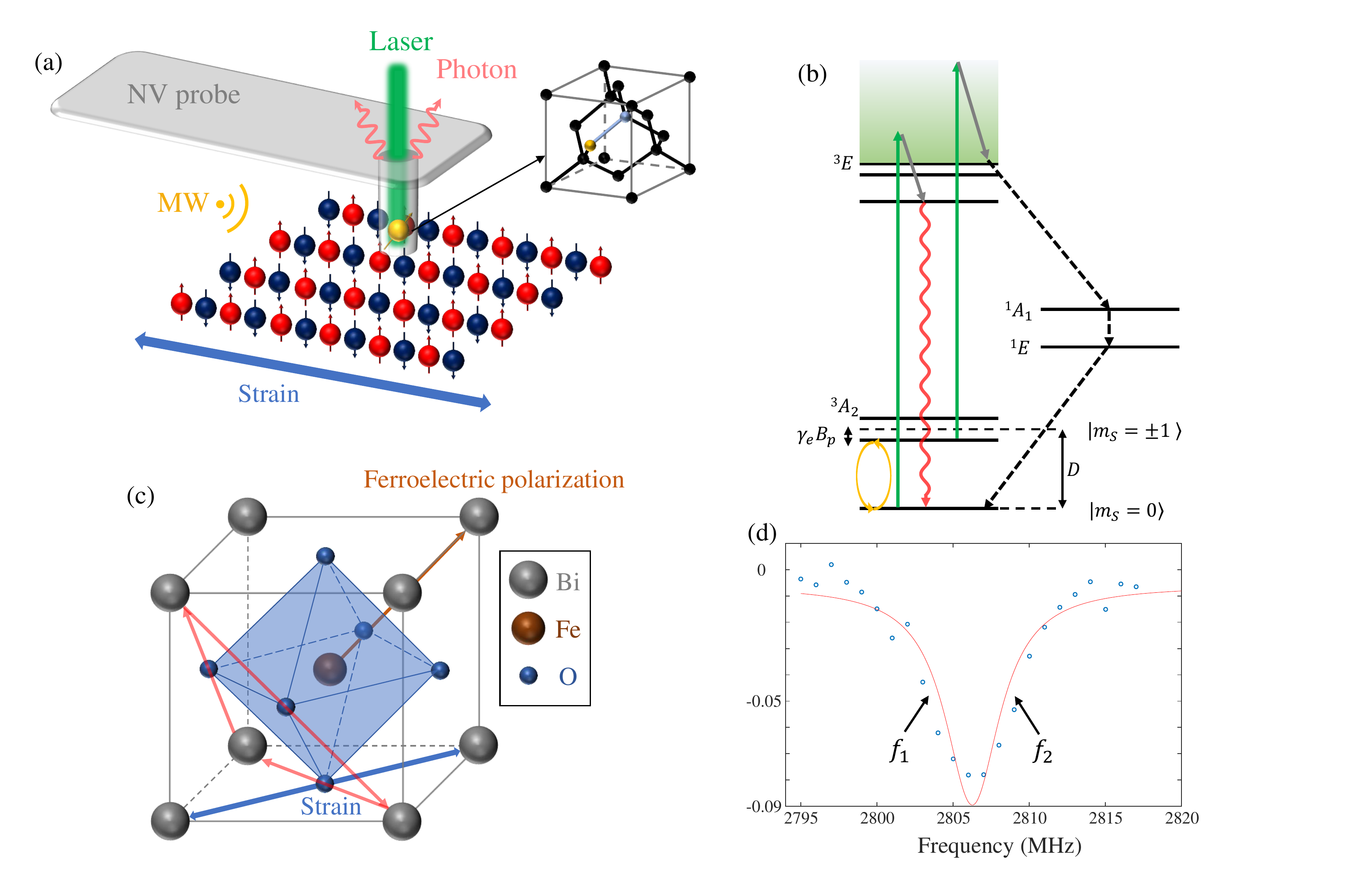} 
\caption{Principle of the experiment. (a) The sketch of an SNVM setup. An NV is at the tip of the probe while its structure is shown in the inset. Under the NV probe, blue and red ball array stands for the antiferromagnetic material imaged during the experiment. Continuous uniaxial strain (blue two-way arrow) is applied \emph{in situ} with an organic substrate (not shown in the sketch).  (b) The energy level structure of an NV center.    (c) The structure of a BFO pseudo-cubic unit cell. The ferroelectric state has been polarized along $[111]$ beforehand (brown arrow) and the uniaxial strain (blue two-way arrow) is applied along $[110]$, which is paralleled to the polarization's projection to the plane. Possible cycloidal wave vectors (light red arrows) in bulk BFO are along face diagonals perpendicular to ferroelectric polarization. (d) The demonstration of dual-iso-B protocol. CW-ODMR data are collected at two frequencies ($f_{1,2}$) \cite{supp}. } 
\label{fig1}
\end{figure*}

BFO is a kind of perovskite compound while fulfills noncentrosymmetric rhombohedral $R3c$ space group\cite{burnsExperimentalistGuideCycloid2020}. The structure of BFO is shown in figure \ref{fig1}(c), for the sake of brevity, we adopt pseudo-cubic unit cell. In each unit cell, bismuth atoms are at eight corners, while at the center lays the iron atom contained by an octahedron constructed by six oxygen atoms. The ferroelectric Curie temperature of BFO is $1103\  \mathrm{K}$\cite{kaczmarekDifferentialThermalAnalysis1975, neatonFirstprinciplesStudySpontaneous2005}, below which the ferroelectric polarization is as high as 100 $\mu$C/cm$^2$ in high quality grown films \cite{wangEpitaxialBiFeO3Multiferroic2003, jiFreestandingCrystallineOxide2019}. The antiferromagnetic order of BFO is characterized as G-type, with N\'{e}el temperature $T_\mathrm{N}=683\ \mathrm{K}$ \cite{kiselevDetectionMagneticOrder1963}. 3d electrons of  Fe$^{3+}$ are the origin of magnetism of BFO, while ferroelectricity and antiferrodistortive break the spatial inversion symmetry which gives rise to a DMI. This interaction leads to a small canting angle between neighbor spins and this produces an effective spin density as large as 0.02 $\mu_B$ per unit cell \cite{albrechtFerromagnetismMultiferroicTextBiFeO2010, grossRealspaceImagingNoncollinear2017}. Under proper conditions, spin distribution in BFO will turn cycloidal, which is an incommensurable periodic order. A cycloidal order can be described by a wave vector $\mathbf k$, as indicated in figure \ref{fig1}(c). The magnetic order of BFO is decided by external field, strain, temperature, size and more factors, while the effect of strain has been intensively  studied\cite{sandoCraftingMagnonicSpintronic2013, haykalAntiferromagneticTexturesBiFeO2020, waterfieldpriceStrainEngineeringMultiferroic2019, sandoMagneticPhaseDiagram2019, sandoInfluenceFlexoelectricitySpin2019, chenComplexStrainEvolution2018}. Previous works utilized different substrates to adjust the epitaxial strain in BFO and found two types of cycloidal order\cite{haykalAntiferromagneticTexturesBiFeO2020, sandoCraftingMagnonicSpintronic2013}. While these researches provide the phase diagram of BFO magnetic order with respect to epitaxial strain, they are not able to impose adjustable strain to BFO \emph{in situ} and this leaves the mechanism of BFO magnetic transformation at critical point an outstanding open question \cite{sandoCraftingMagnonicSpintronic2013, haykalAntiferromagneticTexturesBiFeO2020, burnsExperimentalistGuideCycloid2020}. Besides, previous researches mainly focus on biaxial strain while real-space imaging of magnetic structure under uniaxial strain has not yet been performed. 

In this work, we adopt a new method base on molecular beam epitaxy (MBE) to prepare freestanding BFO film\cite{jiFreestandingCrystallineOxide2019, luSynthesisFreestandingSinglecrystal2016}. A 75-unit-cell-thick BFO (001) film is prepared and transferred to organic substrate Polyethylenenaphthalate(PEN) while epoxy is used as the glue to conduct strain to the BFO film. During experiments, \textbf{uniaxial, continuous and \emph{in situ}} strain is imposed on the BFO film by means of mechanically stretching the PEN substrate \cite{zangGiantThermalTransport, hanGiantUniaxialStrain2020}. In principle, this method is able to impose arbitrary in-plane tensile strain on the film, while in this work the strain principal axis deviates from $[110]$ by $\sim 4.7^\circ$. Such strain breaks the intrinsic $R3c$ symmetry of BFO and provide a way to tune the spin cycloid's direction continuously.  A home-built SNVM is used to perform nanoscale magnetic imaging of its stray field. By using this method, we find that the direction of the cycloidal order is modulated by the uniaxial strain which confirms to a first principle calculation. This phenomenon may help people understanding the transition mechanism of magnetic order under strain\cite{sandoCraftingMagnonicSpintronic2013, burnsExperimentalistGuideCycloid2020} whilst the freestanding film based method can be used in strain-based spintronics, new  heterostructure devices and other new multifunctional devices \cite{jiFreestandingCrystallineOxide2019, hanGiantUniaxialStrain2020, zangGiantThermalTransport}. Our new freestanding film based method in combination with SNVM real-space imaging ability paves a new way to study strain-magnetism coupling in antiferromagnetic materials.

\begin{figure*} 
\centering
\includegraphics[width=0.8\linewidth]{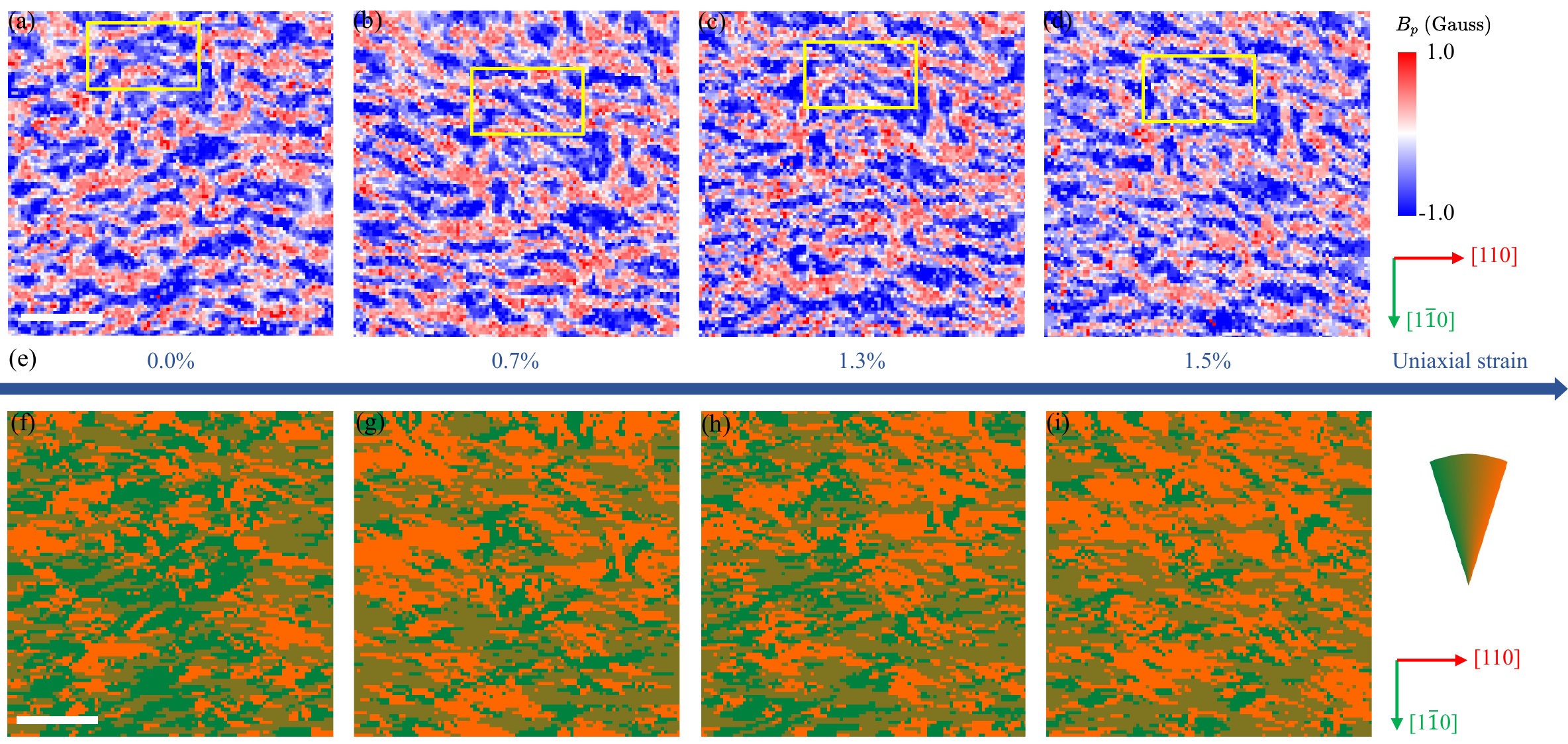} 
\caption{Spin cycloid variation in real-space during the increasing of strain. (a-d) Images of stray field $B_p$ under different strains. The white scale bar corresponds to 500 nm, a direction reference of $[1\overline 1 0]$ and $[110]$ is displayed to the right. Yellow boxes highlight a local transition of the spin cycloid. (e) The sketch of  applied strains corresponding to stray field images above. (f-i) Wave vector direction in real-space under different strains. The white scale bar corresponds to 500 nm. The fan-shaped colorbar demonstrates the wave vector direction under the frame displayed by green and red arrows below. }
\label{fig2}
\end{figure*}

SNVM has been widely used in condensed matter physics\cite{dingzhe2020, casolaProbingCondensedMatter2018, tetienneNanoscaleImagingControl2014, pelliccioneScannedProbeImaging2016, hedrichNanoscaleMechanicsAntiferromagnetic2021, kuImagingViscousFlow2020}, here we apply SNVM to the freestanding BFO film to acquire stray field distribution near the surface. The structure of our SNVM setup is demonstrated in figure  \ref{fig1}(a). The Nitrogen-Vacancy color center (NV center) in diamond is a point defect shown in the inset. It is formed by a nitrogen atom (orange ball in the inset) and an adjacent vacancy (blue ball in the inset) in diamond lattice \cite{dohertyNitrogenvacancyColourCentre2013}. 
As shown in figure \ref{fig1}(b), the NV center is pumped from ground state ($^3 A_2$) into phonon sideband by 532 nm green laser (green arrows) and relaxes into excited state ($^3 E$) with angular momentum conserved (grey arrows).  $|m_\mathrm{S}=0\rangle$ emits photons (wavy red arrow) during the transition to ground state and the photons are finally detected by a single photon detector. $|m_\mathrm{S}=\pm 1\rangle$ evolves into ground state through meta-stable states ($^1A_1, ^1E$) with no detectable photon emitted (black dotted line arrows). The degeneracy of $|m_\mathrm{S}=\pm 1\rangle$ is lifted by Zeeman splitting generated by external magnetic field ($\gamma_e B_p$) and resonant microwave (MW, orange circled arrow) is applied by a copper micro-antenna to selectively excite one of the spin states. By using the photon count rate's difference between $|m_\mathrm{S}=0\rangle$ and $|m_\mathrm{S}=\pm 1\rangle$, it is straight forward to readout the NV center's spin state. 

In our experiment, by applying green laser and MW simultaneously and readout the photon counts, we utilize the Continuous Wave Optically Detected Magnetic Resonance (CW-ODMR) spectrum. In order to accelerate the imaging speed, we adopt the dual-iso-B protocol shown in figure \ref{fig1}(d) \cite{grossRealspaceImagingNoncollinear2017}. Applying this protocol, by sampling at two MW frequencies, we are able to calculate the projection of the stray field to the NV axis\cite{supp}. By scanning across a magnetic film edge, we determine that the distance from the NV center to sample surfaces is 78.5$\pm$1.8 nm (with 95\% confidence) \cite{jenkinsSinglespinSensingDomainwall2019, tetienneNatureDomainWalls2015, supp}.

Beforehand, a piezoelectric force microscope (PFM) is utilized to electrically polarize an area of the BFO film to $[111]$\cite{supp}. During the experiment, a uniaxial strain is applied to the BFO film via the PEN substrate. 
By employing X-ray diffraction (XRD) after the experiment, we are able to calibrate strains under which the images are acquired \cite{supp}. 

SNVM imaging is implemented under four different strains: $\epsilon = 0.0 \%, \ 0.7 \%, \ 1.3\%, \ 1.5 \%$, the results are shown in figure \ref{fig2}. The principal axis of strain deviates from high symmetry direction $[110]$ by 4.7$^\circ$, which breaks $R3c$ symmetry and leads to cycloid tilting.  From the real-space imaging one can find that although the coherence length is relatively small, the sample does possess local cycloidal order, which is modulated by the increasing uniaxial strain. 
We attribute this small coherence length to in-homogeneous strain gradient introduced by epitaxial interface \cite{sandoInterfacialStrainGradients2020}. Be aware that we are using a novel, organic and soft substrate in combination with a freestanding film to realize \emph{in situ} strain tuning.  Without rigid constraint from crystal substrates, freestanding films possess intrinsic unevenness, which may also lead to the small coherence length \cite{zangGiantThermalTransport}. Despite the relatively small coherence length, spin cycloid's variation during the increasing of strain is rather distinct. We calculate the direction of cycloidal wave vector by minimizing variances on segments parallel to different directions and apply region growing algorithm to the results to acquire wave vector direction domain image plotted in figure \ref{fig2}(f-i). It is evident that during the application of strain, the wave vector tilts away from $[1\overline 1 0]$. 

Be aware that the applied strain also modulates local distribution of the magnetic order, there are plural non-trivial local transitions while the applied strain increases. For example, by retracing with respect to tomography markers and magnetic patterns, we are able to determine that the boxed areas in figure \ref{fig2} are at the same spot\cite{area}. Two different ordered areas at this spot merge into one when strain approaches 1.5\%.  This phenomenon is interpreted as the release of local strain gradient under external strain. 

\begin{figure} 
\centering
\includegraphics[width=0.9\linewidth]{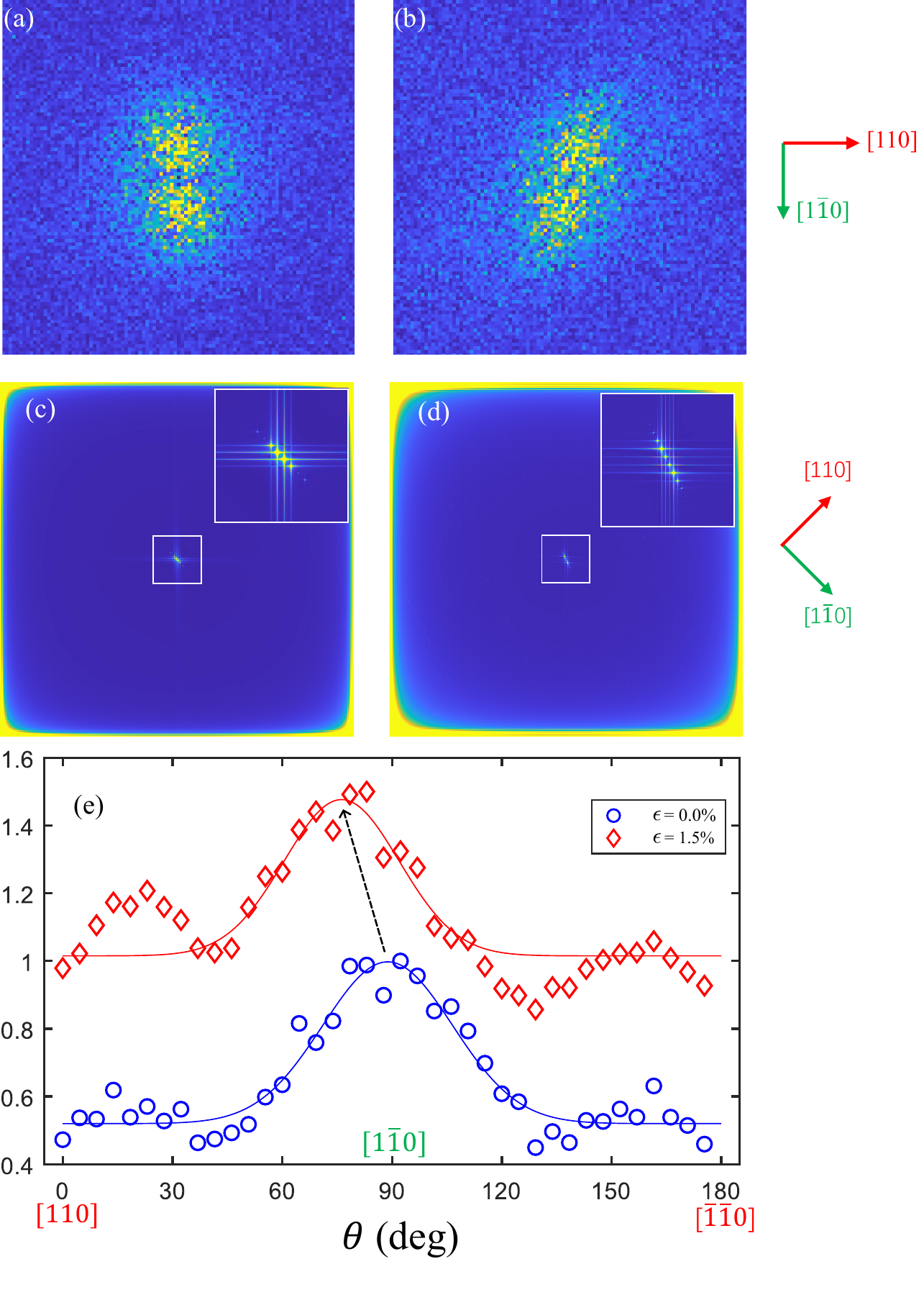} 
\caption{Reciprocal space analysis. (a,b) The FT results of real-space stray field image. (a) corresponds to stray field under no strain and (b) under 1.5\% strain. The reference frame is displayed with green and red arrows to the right. (c,d) The FT results of first principle calculation. (c) is calculated with no strain applied and (d) a 1.5\% uniaxial strain along a direction deviates from pseudo-cubic $[110]$ by 4.7$^\circ$. The insets are enlarged picture of the boxed area close to origin for the sake of clarity of details. The reference frame of the calculation is plotted to the right. (e) Counted results of the distribution of the cycloidal wave vector under different strains with respect to azimuth angle. High lattice orientations are labelled at corresponding azimuth angles. Blue circles and red diamonds stands for data under 0.0\% and 1.5\% strain respectively. Gaussian fit for both peaks are displayed with solid lines according to which there is a tilting angle as large as 12.6$^\circ$. } 
\label{fig3} 
\end{figure}

We have acquired a qualitative result that during the application of strain, the wave vector tilts away from $[1\overline 1 0]$. In order to obtain quantitative relation between the tilting angle and strain, we apply Fourier transformation (FT) to the initial and final real-space scanning results to obtain the wave vector distribution in reciprocal space, which is shown in figure \ref{fig3}(a,b). It is distinctive that the wave vector tilts away from $[1\overline 1 0]$ while the strain increases, this confirms to our qualitative result from the real-space imaging. We accounts the wave vector with respect to its azimuth angle, the result is displayed in figure \ref{fig3}(e). One can find that, while the strain increases, the wave vector tilts by $\sim 12.6^\circ$.  


\begin{figure}
\centering
\includegraphics[width=0.9\linewidth]{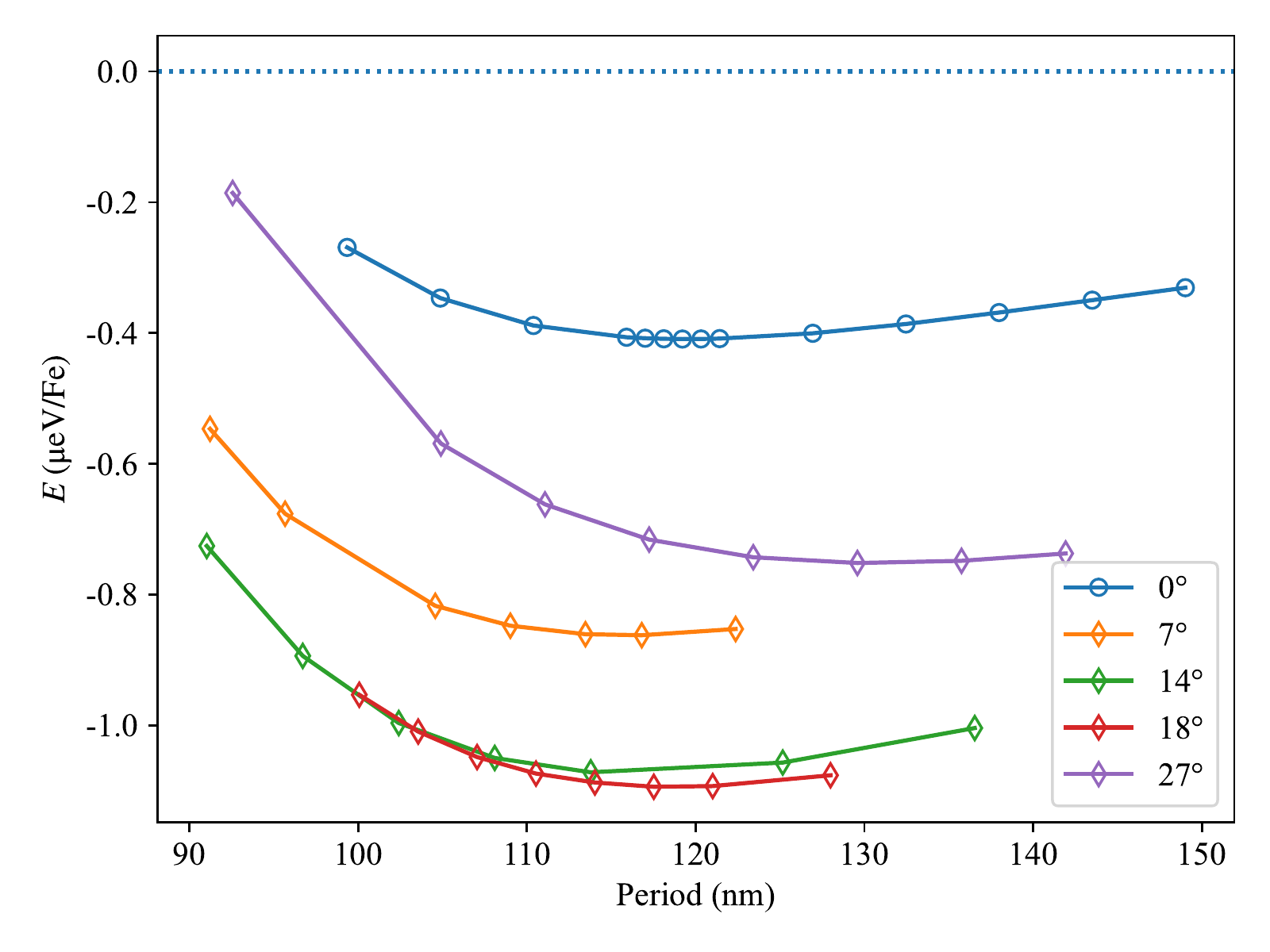} 
\caption{Energy curves with respect to cycloidal period. Data of cycloidal order with wave vector parallel to $[1\overline 1 0]$ are shown with blue circles. Data with wave vectors along different directions ($7^\circ, 14^\circ, 18^\circ, 27^\circ$ away from $[1\overline 1 0]$) are shown with diamonds in different colors. The values are given by the energy differences with respect to the antiferromagnetic state, which is denoted by the horizontal dotted lines. The solid lines between symbols are only guided to eyes.} 
\label{fig4}
\end{figure}

In order to explain the wave vector's tilting under strain, first-principle-based simulations are performed using magnetic effective Hamiltonian \cite{Heff_PhysRevB.99.104420, supp} with the magnetic exchange interaction coefficients evaluated from the four-state mapping approach \cite{FourStates_C2DT31662E}.
 
The uniaxial strain is modeled by stretching the cell along the direction slightly deviates from pseudo-cubic [110] direction by 4.7 $^\circ$. The strain is chosen to be 1.5\% according to XRD calibration, and an estimated Poisson's ratio of about 0.24 calculated via XRD data is also considered \cite{supp}. Note that due to the symmetry broken caused by strain, the symmetry of the system is reduced from space group $R3c$ to $P1$, and all the sixfold-degenerate pair in the $R3c$ structure\cite{Heff_PhysRevB.99.104420} are no longer degenerate.  The calculated exchange coefficients are shown in the Supplemental Information \cite{supp}.

Parallel tempering Monte Carlo (PTMC) simulations \cite{PTMC_B509983H,PASP_doi:10.1063/5.0043703} are performed to solve the magnetic effective Hamiltonian.  A large supercell of $\sqrt 2 n \bf a \times \sqrt 2\bf b \times \bf c$ $ (n=2,3,\cdots, 300)$ is used study the cycloidal phase, where {\bf a} is along the pseudo-cubic $[1\bar 10]$ direction, {\bf b} is along the pseudo-cubic $[110]$ direction, and {\bf c} is along the pseudo-cubic $[001]$ direction. To study the stable period of the cycloidal phase, energy per Fe atom is calculated as a function of cycloidal period, as shown in Fig.  \ref{fig4}. This energy-versus-period curve shows a minimum at 119 nm, which is similar to that in the $R3c$ phase \cite{Heff_PhysRevB.99.104420}.

To study cycloid which deviates the $[1\overline 1 0]$ direction, several \emph{oblique} supercells are constructed. The supercells slightly tilting about {\bf c} direction were used. The angle between the  lattice vector $\mathbf a$ and the pseudo-cubic $[1\overline{1}0]$ direction is used to define the oblique angle of cycloid. 
For example, one of the oblique supercells used in this work is defined by $\mathbf{a} = 211\mathbf{a}_0 - 165\mathbf{b}_0, \mathbf{b}=7\mathbf{a}_0+9\mathbf{b}_0, \mathbf c=\mathbf c_0$ where $\mathbf a_0, \mathbf b_0$ and $\mathbf c_0$ are the vectors we used to define our relaxed $R3c$ BFO lattices which are close to the pseudo-cubic $[100], [010]$ and $[001]$ direction, respectively. The oblique angle of such cell is then the angle between the $\mathbf a$ and $\mathbf{a_0}-\mathbf{b_0}$, that is approximately 7$^\circ$. Figure \ref{fig4} shows the calculated energy of supercells with different oblique angles as functions of cycloidal period. The energies of supercells with non-zero oblique angles are lower than that of supercells  without oblique angle, indicating the oblique supercells are more energetically favorable. Moreover, 
with the increasing of the oblique angle from 0$^\circ$, the energy shows a nonmonotonic behavior with respect to the oblique angle, with a minimum around 18$^\circ$. Such value is in  in very well consistent with the measured value 12.6$^\circ$ (refer to figure \ref{fig3}(e)). 

To confirm the cycloids in the rotated supercells do propagate along a direction deviates from $[1\bar 10]$, FT is also performed to the PTMC simulation results \cite{supp}.
Figure \ref{fig3} (d) shows the spectrum of one supercell with oblique angle 18 $^\circ$ after the FT, showing a peak at the coordinate $(12,-6)$, which is away from the pseudo-cubic $[1\overline 10]$ direction by 18$^\circ$ (calculated from $45^\circ - \arctan (6/12)$). On the other hand, figure  \ref{fig3} (c) shows such spectrum on a supercell without oblique angle of non-stretched cell ($R3c$ phase as in Ref. \cite{Heff_PhysRevB.99.104420}), showing a peak at the coordinate $(10,-10)$, which is exactly along the pseudo-cubic $[1\overline 10]$ direction. The deviation of such peak from pseudo-cubic $[1\overline 10]$ direction definitely confirms that the cycloids in oblique cells do propagate along a direction deviates from $[1\overline 10]$ direction. We thus conclude that such deviation of the cycloid is energetically more favorable that without deviation.

Note that we also performed  above analysis on BFO at tensile strain of 0.5\% along \emph{exactly} the pseudo-cubic [110] direction without oblique angle, and found the cycloid propagates along the pseudo-cubic $[1\overline{1}0]$ (i.e. with oblique angle 0$^\circ$) is the most energetically favored phase. Such fact indicates that the deviation of the cycloid propagation originates from the symmetry broken caused by the uniaxial strain {color{red} with an oblique angle}.

In summary, we apply contineous \emph{in situ} uniaxial strain to freestanding BFO films and scan the magnetic stray field near surface with SNVM. A modulation of direction of the magnetic order is found and a first principle calculation is conducted. The first principle calculation result confirms that the strain induced magnetic effective Hamiltonian evolution is responsible for the rotation of magnetic order. This result is helpful for people to understand the mechanism of magnetic order transformation under strains. Besides, our freestanding-film-based \emph{in situ} strain applying method paves a new way to study the coupling between strain and magnetism in antiferromagnetic materials. With some slight improvements, it is possible to conduct arbitrary in-plane tensile strain to the sample, which provides people with higher degrees of freedom to study the mechanism of magnetic order transition under strains.


\begin{acknowledgments}
This work was supported by the National Natural Science Foundation of China (grant nos. 81788101, T2125011,  No.12104447), the National Key R\&D Program of China (grant nos. 2018YFA0306600 and 2021YFA1400400), the CAS (grant nos. XDC07000000, GJJSTD20200001, QYZDY-SSW-SLH004, Y201984, YSBR-068), Innovation Program for Quantum Science and Technology (Grant No. 2021ZD0302200, 2021ZD0303204), the Anhui Initiative in Quantum Information Technologies (grant no. AHY050000), the Natural Science Foundation of Jiangsu Province (grant No. BK20200262), Hefei Comprehensive National Science Center, China Postdoctoral Science Foundation (Grant No. 2020M671858) and the Fundamental Research Funds for the Central Universities.

We are grateful to the HPCC resources of Nanjing University for the calculations. The NV scanning probe was provided by CIQTEK and the assembling was completed by Ruige Shao. This work was partially carried out at the USTC Center for Micro and Nanoscale Research and Fabrication.
 
\end{acknowledgments}

\bibliography{BFObib.bib}

\begin{thebibliography}{10}
\bibitem{area}
 \bibinfo{pages}{When we adjust strain, a movement at hundred micrometer scale
  is introduced by the strain applying positioner. A built--in optical
  microscope supplies tomography information to help reset the sample's
  position. The limited spatial resolution of the optical microscope is the
  reason that the field of views under different strains slightly mismatch,
  though we could retrace common areas by referring to neighbor magnetic
  patterns.}
\end{thebibliography}
\end{document}


\title{Supplemental information \\ Observation of uniaxial strain tuned spin cycloid in a freestanding BiFeO$_3$ film}
\maketitle
\author{A. Author}
\section{Reconstructing stray field from dual-iso-B data}
In the main text, we mentioned that CW-ODMR is utilized to acquire the stray field imaging near BFO surface, this section explains the principle of CW-ODMR and dual-iso-B protocol. 

In this paragraph, while analyzing NV electron ODMR, we utilize NV local frame. NV axis is defined as z-axis and spin states along z-axis are taken as basis: $\{|+1\rangle_e, |0\rangle_e, |-1\rangle_e \}$. The operator of electron spin's z-component is: 
\begin{equation}
S_z = \begin{pmatrix}
 1 & 0 & 0\\
 0 & 0 & 0\\
 0 & 0 & -1
\end{pmatrix}. 
\end{equation}
NV centers utilized in the experiment are prepared by $^{15} \rm N$ ion implanting followed by annealing under 900 $^\circ$C. Noting that an $^{15} \rm N$'s nuclear spin quantum number is 1/2, by taking spin states along z direction $\{|+1/2\rangle_n, |-1/2\rangle_n \}$ as basis, the operator of nuclear spin's z-component is
\begin{equation}
I_z = \frac{1}{2}\ \begin{pmatrix}
 1 & 0 \\
 0 & -1 \\
\end{pmatrix}.
\end{equation}
Consider the Hilbert space of NV electron and nuclear spin, we could write the Hamiltonian under secular approximation:  
\begin{equation}
	H_{\rm NV}=D S_z^2-\gamma_e B_z S_z + A_{\rm N} S_z I_z .
\end{equation}

In this formula, $B_z$ is the external magnetic field's projection on z direction, while $D=2870\ {\rm MHz}$ is the zero-field splitting; $\gamma_e = 2.8\ \rm MHz/Gauss$ is the gyromagnetic ration of electron; $A_{\rm N} = 3.03\ \rm MHz$ is $^{15}\rm N$'s hyperfine coupling strength. During the experiment, we only collected the data of resonance between $|+1\rangle_e$ and $|0\rangle_e$. By calculating the eigenvalues of the Hamiltonian, one can find that with respect to the nuclear spin states, there are two resonant frequiencies: $f_\pm=D-\gamma_e B_z \pm A_{\rm N}/2$. In the CW experiment, for each frequency, two sets of data are collected: one set is the photon count $C(f)$ under MW and pump laser; the other set is the photon count $C_0$ under pump laser without MW. By taking these two sets of data, we define the CW spectrum: $S(f) := C(f)/C_0-1$. It has been proven that this spectrum is Lorentzian\cite{dreauAvoidingPowerBroadening2011}. The nuclear spin is under mixed state $\rho_n = 1/2 (|+1/2\rangle_n  \langle +1/2|+|-1/2\rangle_n  \langle -1/2|)$ so the spectrum is the superposition of two Lorentzian lines: 
\begin{eqnarray}
S(f,B_z)= &-& \left[ \frac { A } { ( f - f_+(B_z) ) ^ { 2 } / \Delta ^ { 2 } + 1 }+ \right. \\ \nonumber
& & \left. \frac { A } { ( f - f_-(B_z) ) ^ { 2 } / \Delta ^ { 2 } + 1 }\right],
\end{eqnarray}
in which $A, \Delta$ are contrast and half width at half maximum of one Lorentzian line and calibrated before the experiment. 

Under the dual-iso-B mode, we only collect data at two frequency points $S_i=S(f_i,B_z), i=1,2$, in order to acquire a relatively high sensitivity, we take $f_1=f_--\Delta,\ f_2=f_++\Delta$. Notice that experiment data at each frequency point provides a quartic algebraic equation with respect to $B_z$, we can locate the solution $B_z$ by comparing solution sets under two frequency points. It is necessary to point out that in order to lift the degeneracy between $|\pm 1\rangle_e$, external magnetic field $\mathbf B_{\rm ext}$ is applied and the solved $B_z$ above is the superposition of the external magnetic field and stray field $B_z = B_{{\rm ext},z}+B_{{\rm str},z}$. Since $\mathbf B_{\rm ext}$ has been calibrated beforehand, it is straightforward to acquire the stray field by subtracting it from the solved magnetic field. 

\section{Calibrating NV height}
\begin{figure*}
\centering
\includegraphics[width=0.8\linewidth]{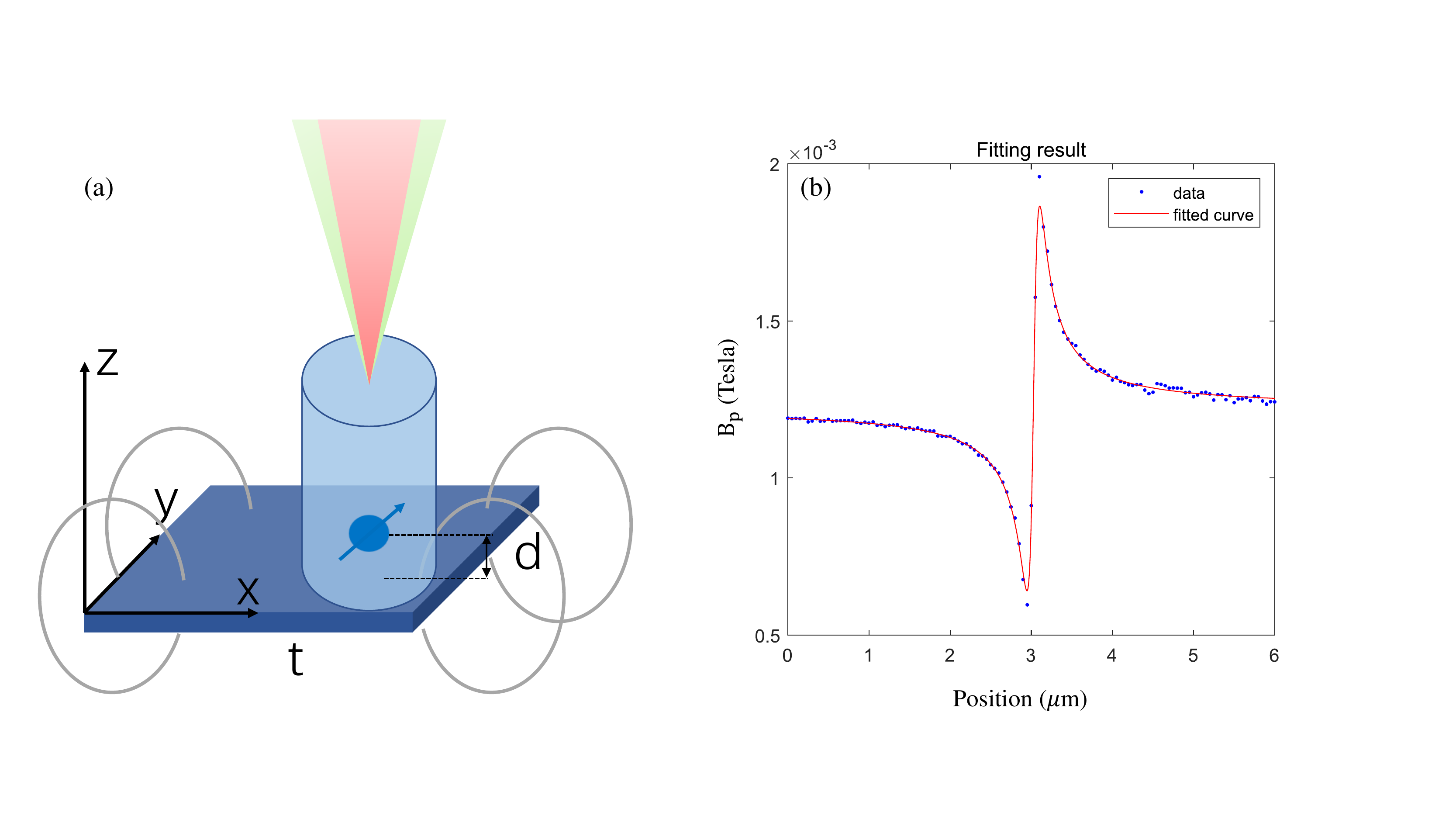} 
\caption{Calculating NV height.(a) is the layout of the experiment for the height calculation. (b) is the line scan and fitting results.  } 
\label{nvheight}
\end{figure*}

By scanning across ferromagnetic film's edges, we are able to acquire the information about the height of the NV center\cite{tetienneNatureDomainWalls2015, jenkinsSinglespinSensingDomainwall2019}. Under the reference frame shown in FIG.\ref{nvheight}(a), the stray field of a ferromagnetic film magnetized along z axis distributes according to the formula below: 
\begin{eqnarray}
B_{x}&=&\frac{\mu_{0} M_{s} t}{2 \pi} \frac{h}{\left(x-x_{0}\right)^{2}+h^{2}}, \\
B_{z}&=&-\frac{\mu_{0} M_{s} t}{2 \pi} \frac{x}{\left(x-x_{0}\right)^{2}+h^{2}}.
\end{eqnarray}
In this formula, $h$ is the height of NV, $M_s, t$ are saturation magnetization and thickness of the ferromagnetic film respectively. $x_0$ is the position of the edge of the film and  $\mu_0$ is the vacuum permeability. Since the experiment data is $B_p=\mathbf B \cdot \mathbf n_{\rm NV}$, by applying the formula above, one can calculate $h$ by fitting $B_p$. 

We proceed line scan across the edge of ferromagnetic film Ta(5 nm)/CoFeB(1 nm)/MgO(1.2 nm)/Ta(2 nm) and fit the $B_p$ data to calculate the NV height. The data is shown in  FIG.\ref{nvheight}(b) and the fitting results shows that the NV height is 78.5$\pm$1.8 nm (with 95\% confidence). 

\section{Electric polarization of BFO}
\begin{figure*}
\centering
\includegraphics[width=0.9\linewidth]{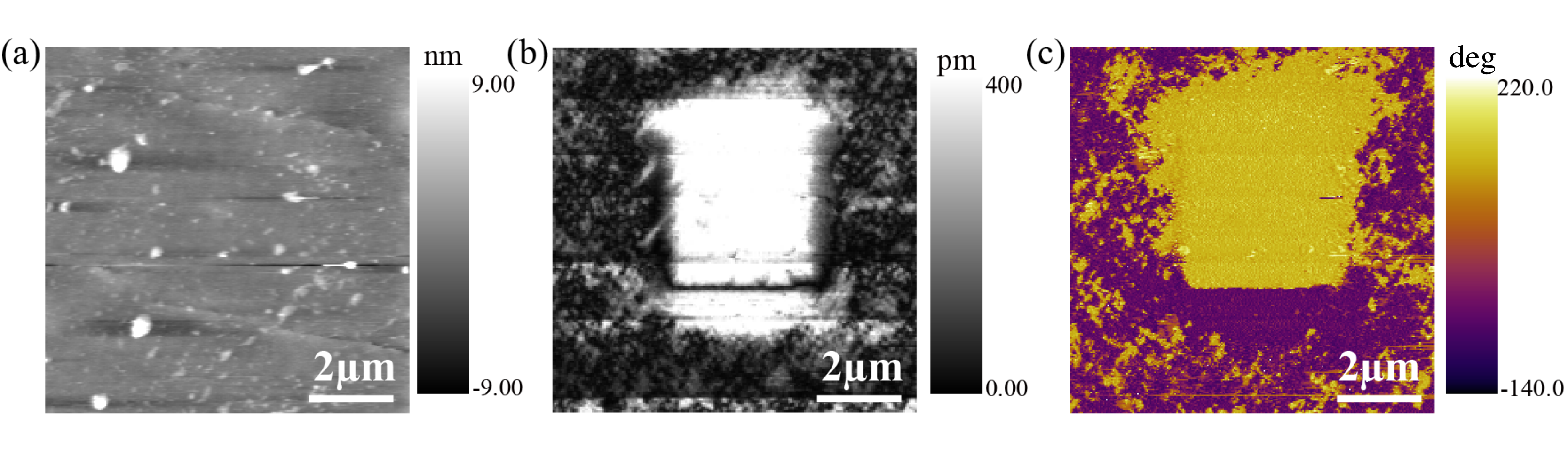} 
\caption{PFM results. (a), (b) and (c) are topography, amplitude and phase data respectively. } 
\label{pfm}
\end{figure*}

Piezoelectric force microscopy (PFM) has been utilized to electrically polarize the BFO film along $[111]$ before experiments and LPFM has been used to confirm that BFO has indeed been polarized. Results of the PFM are shown in FIG.\ref{pfm}. 

\section{Strain calibration and structure analysis}
\begin{figure*}
\centering
\includegraphics[width=0.9\linewidth]{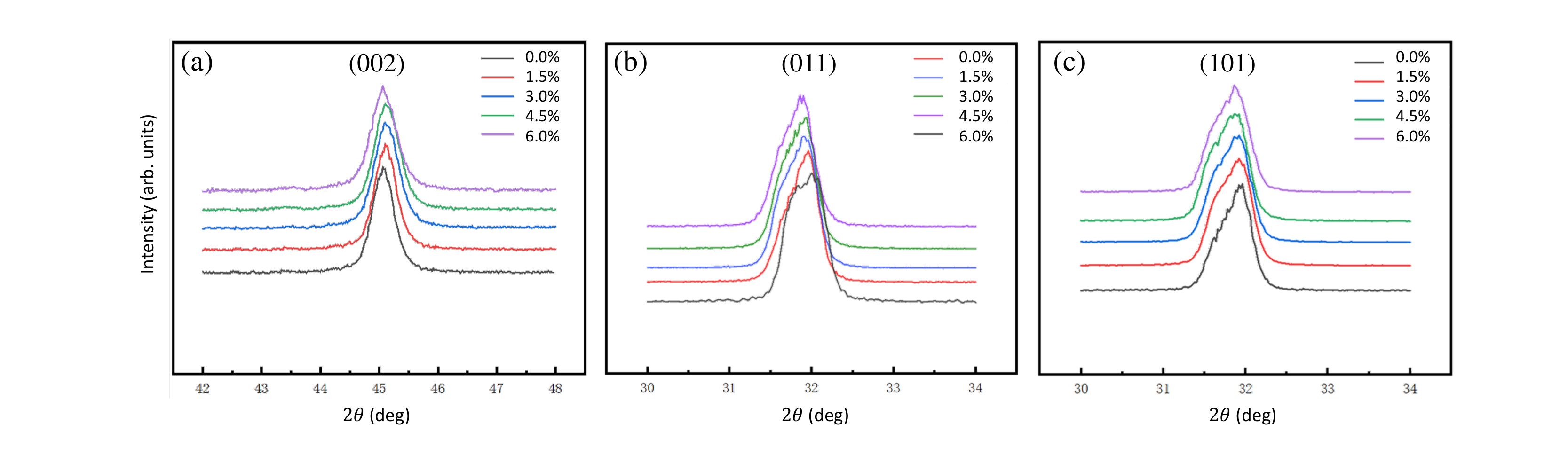} 
\caption{2$\theta$-$\omega$ scans of the BFO films under different nominal strain. (a),(b),(c) are diffraction patterns of $(002)$, $(011)$ and $(101)$ planes respectively.} 
\label{2thetaData}
\end{figure*}

Strains under which the experiments have been performed are calibrated by X-ray diffraction (XRD) afterwards. During experiments, stretching lengths of the substrate $\Delta L$ are measured by the strain applying positioner's sensor and a nominal strain $\varepsilon_n = \Delta L/L\times 100\%$ is defined. Here, $L$ is the length of the substrate. SNVM scanning in the main text are performed under $\varepsilon_n = 0.0,\ 1.5,\ 3.0,\ 4.5\%$.  The 2$\theta$-$\omega$ scans under different nominal strains are displayed in FIG.\ref{2thetaData}. 

\begin{figure*}
\centering
\includegraphics[width=0.9\linewidth]{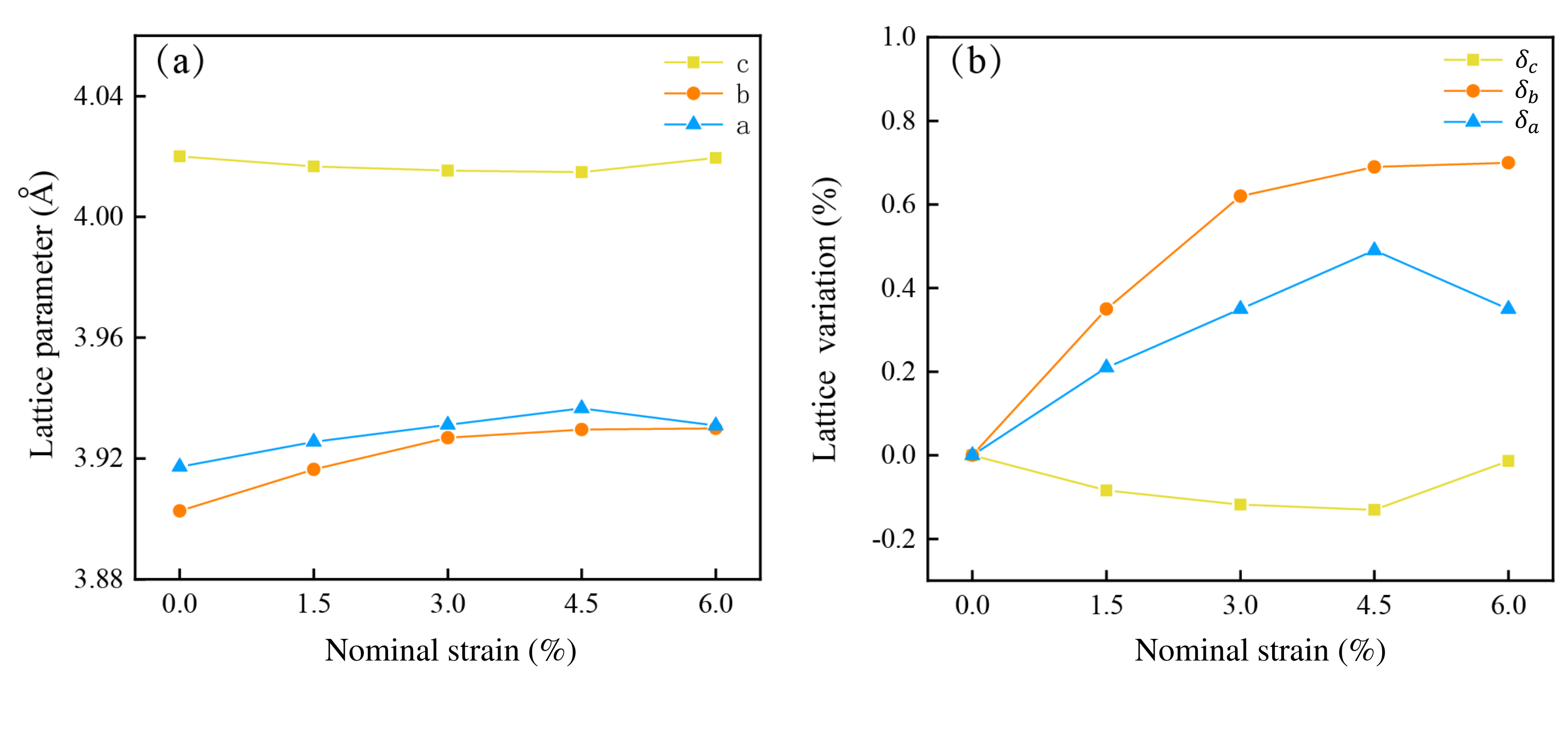} 
\caption{(a) is the relation between lattice parameters and nominal strain. (b) is the relation between lattice variation and nominal strain. } 
\label{lattice}
\end{figure*}

One can calculate the lattice parameters by fitting the 2$\theta$-$\omega$ data. The relation between lattice parameters $a,b,c$ and nominal strain is shown in FIG.\ref{lattice}(a). We define lattice variations: 
\begin{eqnarray}
\delta_a &=& \frac{a-a_0}{a_0} \times 100\%,  \\
\delta_b &=& \frac{b-b_0}{b_0} \times 100\%,  \\
\delta_c &=& \frac{c-c_0}{c_0} \times 100\%.  
\end{eqnarray}
Here, $a_0, b_0, c_0$ are the lattice parameters under zero nominal strain.  There is a inflection point after nominal strain approaches 4.5\%. After that point, microcracks appears so that the strain starts to relax \cite{zangGiantThermalTransport}. 

For brevity, we only consider the in-plane strain and the strain tensor under principal axis coordinate is 
\begin{equation}
    \begin{pmatrix}
  \epsilon & 0 \\
  0 & \sigma \epsilon
\end{pmatrix}
\end{equation}
in which $\epsilon$ is the strain component under the stretching direction (corresponding to strain mentioned in the main text) and $\sigma$ represents for Poisson's ratio. We transform the coordinate so that the first axis is parallel to lattice vector $\mathbf a$
\begin{equation}
\begin{pmatrix}
  \cos\theta_a & -\sin\theta_a \\
  \sin\theta_a & \cos\theta_a
\end{pmatrix} 
\begin{pmatrix}
  \epsilon & 0 \\
  0 & \sigma \epsilon
\end{pmatrix} 
\begin{pmatrix}
  \cos\theta_a & \sin\theta_a \\
  -\sin\theta_a & \cos\theta_a
\end{pmatrix} = 
    \begin{pmatrix}
  u_{aa} & u_{aa'} \\
  u_{a'a} & u_{a'a'}
\end{pmatrix}.
\end{equation}
Here, $\theta_a$ is the rotation angle of the coordinate transformation, $a'$ labels the direction perpendicular to $\mathbf a$. It can be proven that 
\begin{equation}
    u_{aa} = \delta_a.
\end{equation}
Note that the pseudo-cubic structure is broken by strain, $a'$ is not necessarily parallel lattice vector $\mathbf b$.  By observing the $1,1$-term of the strain tensor,  we could acquire the relation between strains and lattice variation $\delta_a$. By performing a similar transformation so that the first axis is parallel to lattice vector $\mathbf b$, we derive such equations:
\begin{eqnarray}
\epsilon(\cos^2\theta_a+\sigma\sin^2\theta_a)&=&\delta_a, \label{strain_aa}\\
\epsilon(\cos^2\theta_b+\sigma\sin^2\theta_b)&=&\delta_b.\label{strain_bb}
\end{eqnarray}

Since strain changes the angle $\gamma$ between $\mathbf a$ and $\mathbf b$, we shall calculate $\gamma$ from XRD data. In fact, one can acquire this angle by calibrating the change of $\phi$ between $(101)$ and $(011)$ interference patterns. Since the stretching direction does not deviate from $[110]$ much, we could acquire an approximate equation of strain with respect to $\gamma$
\begin{equation}
    \frac{1+\sigma \epsilon}{1+\epsilon}=\frac{\tan\gamma/2}{\tan\gamma_0/2}. \label{gamma}
\end{equation}

\begin{figure*}
\centering
\includegraphics[width=0.9\linewidth]{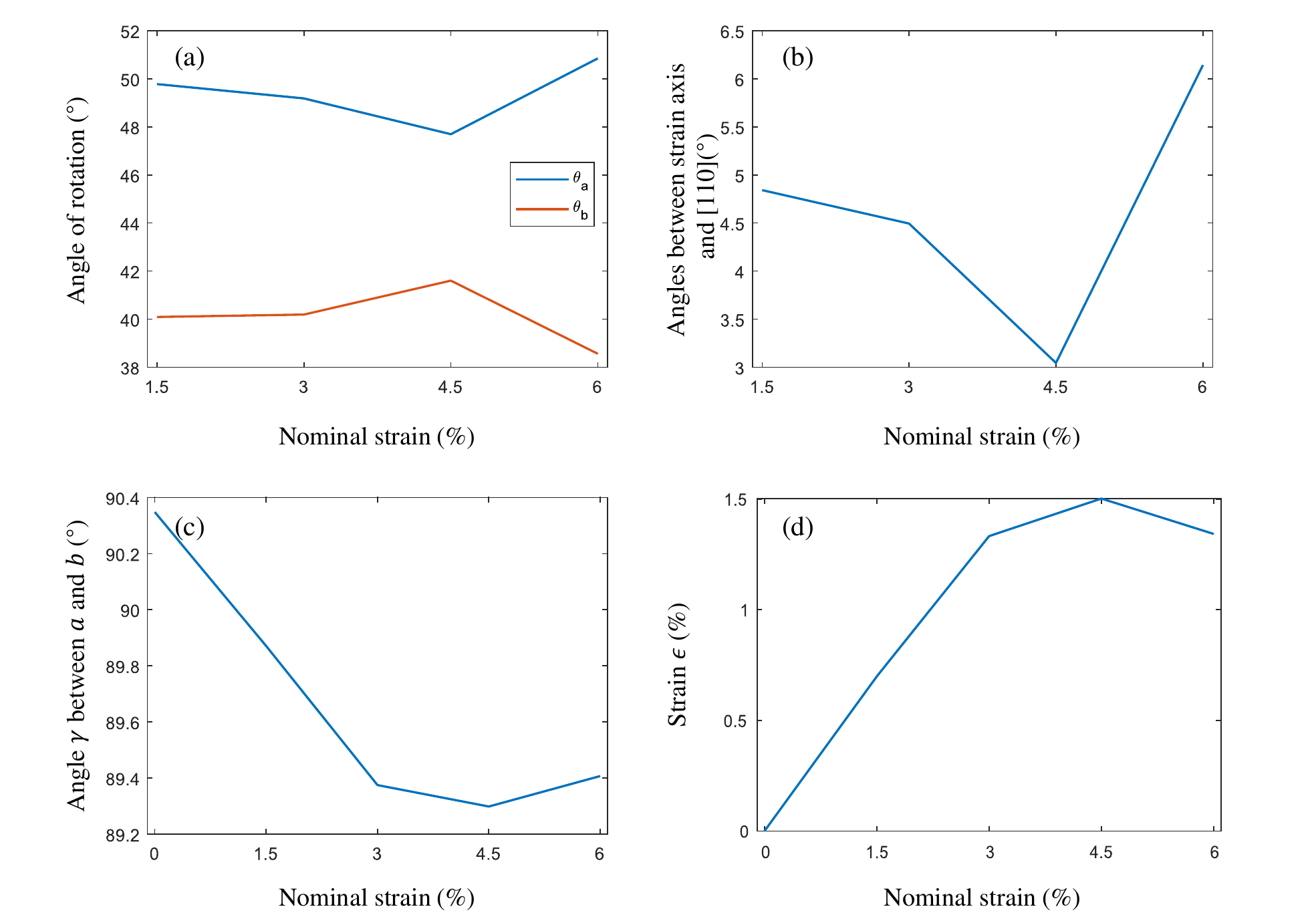} 
\caption{Analysis results from XRD data. (a) Coordinate rotation angles mentioned in equation (\ref{strain_aa}) and (\ref{strain_bb}). (b) Strain deviating angle from high symmetry direction $[110]$. (c) Angle $\gamma$ between lattice vectors $\mathbf a$ and $\mathbf b$ calculated from XRD data. (d) Relation between strain $\epsilon$ and nominal strain calculated from XRD data. } 
\label{solvexrd}
\end{figure*}

Combine equation (\ref{strain_aa}), (\ref{strain_bb}), (\ref{gamma}) and note that geometric relation $\theta_a+\theta_b=\gamma$ provides another constraint, we are able to solve $\theta_a, \theta_b, \epsilon$ and $\sigma$ at the same time (shown in FIG.\ref{solvexrd}). 

\section{First principle calculation results}
In the first principle calculation, we adopt such a magnetic effective Hamiltonian: 
\begin{equation}
E = E_{\rm ex} + E_{\rm dm}+E_{\rm ani}, 
\end{equation}
which are exchange interaction, Dzyaloshinskii-Moriya interaction (DMI) and single ion anisotropy (SIA) respectively. The forms of the three components are: 
\begin{align}
E_{\mathrm{ex}}&=&\frac{1}{2} \sum_{\langle i, j\rangle} \mathbf{S}_{i} \cdot \mathcal{J}_{\langle i,j \rangle} \cdot \mathbf{S}_{j},\\
E_{\mathrm{dm}}&=&\frac{1}{2} \sum_{\langle i,j\rangle}\mathbf D_{\langle i,j \rangle}\cdot(\mathbf S_{i}\times \mathbf S_{j}),\\
E_{\mathrm{ani}}&=&\sum_{i} \mathbf{S}_{i} \cdot \mathcal{A}_{i} \cdot \mathbf{S}_{i}.
\end{align}
Here, $\mathbf{S}_{i}$ is the spin operator at site $i$, $\mathcal J_{\langle i,j \rangle}$ and $\mathbf D_{\langle i,j \rangle}$ are the exchange interaction tensor and DM vector between site $i$ and $j$. $\mathcal A_i$ is the single ion anisotropy at site $i$. In the calculation, we only consider the coupling between nearest neighbors and next nearest neighbors.

As mentioned in the main text, after the coefficients the magnetic effective Hamiltonian is acquired by the four-state mapping approach, it is used in parallel tempering Monte Carlo (PTMC) simulations to study the cycloidal phase. The calculated exchange coefficients and SIA coefficients are shown in Tabs. \ref{tab_exch} and \ref{tab_sia}, respectively.

\begin{table*}[htb]
	\caption{Calculated exchange parameters and DM intereactions (in meV) for Fe-Fe pairs in stretched BFO. The atom positions and distances are given in cartesian coordinates in Angstroms.}
	\label{tab_exch}
	\begin{ruledtabular}
	\begin{tabular}{llllllllll}
		 & atom 1 & atom 2 & distance & $J_{xx}$ & $J_{yy}$ & $J_{zz}$ & 
		 $D_x$ & $D_y$ & $D_z$ \\
		 \hline
		 1 & (6.10, 5.93, 1.95) & (6.13, 5.96, 5.86) & 3.91 & 6.152 & 6.151 & 6.138 & 0.029 & -0.120 & -0.044 \\
		 2 & (2.17, 5.96, 5.87) & (2.14, 5.93, 1.95) & 3.91 & 6.136 & 6.137 & 6.122 & -0.119 & 0.031 & -0.041 \\
		 3 & (6.10, 5.93, 1.95) & (6.00, 1.99, 1.95) & 3.95 & 5.876 & 5.863 & 5.875 & 0.022 & -0.039 & -0.101 \\
		 4 & (6.03, 2.01, 5.87) & (6.13, 5.96, 5.86) & 3.95 & 5.910 & 5.899 & 5.910 & -0.104 & -0.044 & 0.027 \\
		 5 & (2.17, 5.96, 5.87) & (6.13, 5.96, 5.86) & 3.96 & 5.776 & 5.788 & 5.789 & -0.038 & 0.023 & -0.097 \\
		 6 & (6.10, 5.93, 1.95) & (2.14, 5.93, 1.95) & 3.96 & 5.819 & 5.830 & 5.831 & -0.046 & -0.098 & 0.026 \\
		 7 & (2.17, 5.96, 5.87) & (6.03, 2.01, 5.87) & 5.52 & 0.180 & 0.180 & 0.181 & 0.001 & 0.000 & 0.018 \\
		 8 & (2.14, 5.93, 1.95) & (6.00, 1.99, 1.95) & 5.52 & 0.182 & 0.182 & 0.183 & 0.000 & 0.001 & 0.018 \\
		 9 & (2.14, 5.93, 1.95) & (2.07, 2.01, 5.86) & 5.54 & 0.183 & 0.182 & 0.182 & -0.019 & -0.002 & -0.003 \\
		 10 & (6.10, 5.93, 1.95) & (6.03, 2.01, 5.87) & 5.54 & 0.174 & 0.172 & 0.173 & -0.019 & -0.002 & -0.002 \\
		 11 & (6.10, 5.93, 1.95) & (2.17, 5.96, 5.87) & 5.55 & 0.186 & 0.187 & 0.185 & 0.002 & 0.019 & 0.002 \\
		 12 & (6.00, 1.99, 1.95) & (2.07, 2.01, 5.86) & 5.55 & 0.174 & 0.175 & 0.174 & 0.002 & 0.019 & 0.002 \\
		 13 & (6.00, 1.99, 1.95) & (6.13, 5.96, 5.86) & 5.58 & -0.004 & -0.005 & -0.005 & -0.005 & 0.003 & -0.001 \\
		 14 & (2.04, 1.98, 1.95) & (2.17, 5.96, 5.87) & 5.58 & -0.005 & -0.006 & -0.006 & 0.003 & 0.000 & -0.001 \\
		 15 & (2.14, 5.93, 1.95) & (6.13, 5.96, 5.86) & 5.59 & -0.003 & -0.001 & -0.003 & 0.000 & -0.003 & 0.001 \\
		 16 & (2.04, 1.98, 1.95) & (6.03, 2.01, 5.87) & 5.59 & -0.003 & -0.001 & -0.003 & -0.003 & 0.005 & 0.001 \\
		 17 & (2.07, 2.01, 5.86) & (6.13, 5.96, 5.86) & 5.66 & 0.013 & 0.013 & 0.015 & 0.003 & 0.000 & -0.004 \\
		 18 & (2.04, 1.98, 1.95) & (6.10, 5.93, 1.95) & 5.66 & 0.015 & 0.014 & 0.016 & 0.000 & -0.002 & 0.004 \\
				  
	\end{tabular}
	\end{ruledtabular}
\end{table*}

\begin{table*}
	\caption{Calculated SIA parameters (in $\mu$eV) for Fe atoms. The atom positions are given in cartesian coordinates in Angstroms.}
	\label{tab_sia}
	\begin{ruledtabular}
		\begin{tabular}{lllllllll}
			  & atom & $\mathcal{A}_{yy}-\mathcal{A}_{xx}$ & $\mathcal{A}_{zz}-\mathcal{A}_{xx}$ &
			  $\mathcal{A}_{xy}$ & $\mathcal{A}_{xz}$ & $\mathcal{A}_{yz}$ & \\
			  \hline
			  1 & (2.04, 1.98, 1.95) & 0.7 & 0.3 & -2.7 & -0.4 & -3.1 \\
			  2 & (2.14, 5.93, 1.95) & 0.7 & 0.4 & -2.7 & -0.4 & -3.1 \\					
		\end{tabular}
	\end{ruledtabular}
\end{table*}

Practically, it is not possible to perform \emph{discrete} Fourier transformation on the rotated supercell directly, since the lattice vectors $\mathbf a$ and $\mathbf b$ are not perpendicular to each other, and the primitive cells cannot be discretized into uniform mesh with respect to the lattice vectors. 
An approximate FT is thus performed as follows.
The coordinates of the Fe atoms in a $(001)$ plane of the supercell are first transformed into an infinite plane with basis $\mathbf a_0$ and $\mathbf b_0$, where the periodicity along the $\mathbf a$ and $\mathbf b$ supercell lattice vectors is used through the transformation. The discretized coordinates are given within the unit of $\mathbf a_0$ and $\mathbf b_0$ with one of the Fe atoms as the origin. 
An image could then be created with the values of the $x$- (or equivalently, $y$- or $z$-) component of the spins in a (finite)  rectangular range in this plane, to imitate the testing image in the experiments. To reduce the undesirable influence of the periodicity introduced by the image boundary, a relatively large size of $4000 \times 4000$ is used. Then, fast Fourier transformation is performed on this image as usual.

\bibliography{suppBib.bib}